\documentclass[acmsmall]{acmart}

\usepackage[ruled,vlined,lined,commentsnumbered]{algorithm2e}
\usepackage{amsmath,amsfonts}
\usepackage{booktabs}
\usepackage[skip=1pt,labelfont=bf]{caption}
 \usepackage{calligra}
\usepackage{color, colortbl}
\usepackage{courier}
\usepackage{csvsimple}
\usepackage{enumitem}
\usepackage{fancybox}
\usepackage{fontenc}
\usepackage{graphicx}
\usepackage{listings}
\usepackage{longtable}
\usepackage{lscape}
\usepackage{makecell}
\usepackage{marvosym}
\usepackage{moreverb}
\usepackage{multicol}
\usepackage{multirow}
\usepackage{pifont}
\usepackage{rotating}
\usepackage{setspace}
\usepackage{caption}
\usepackage{subfigure}
\usepackage[most]{tcolorbox}
\usepackage{threeparttable}
\usepackage{tikz}
\usepackage[normalem]{ulem}
\usepackage{url}
\usepackage{soul}
\usepackage{wasysym}
\usepackage{svg}
\usepackage{textcomp}
\usepackage{xcolor}
\usepackage{wrapfig}
\usepackage{pdflscape}
\usepackage{hyperref}

\usepackage{amsmath}
\usepackage{array}
\usepackage{balance}
\usepackage{microtype}
\usepackage[skip=1pt,labelfont=bf]{caption}

\newboolean{showcomments}
\setboolean{showcomments}{true}
\ifthenelse{\boolean{showcomments}}
 { \newcommand{\mynote}[2]{
      \fbox{\bfseries\sffamily\scriptsize#1}
        {\small$\blacktriangleright$\textsf{\emph{#2}}$\blacktriangleleft$}}}
        { \newcommand{\mynote}[2]{}}

\newcommand{\code}[1]{{\footnotesize\texttt{#1}}}

\newcommand{\toolname}{Rectifier\xspace}

\newcommand{\intuition}[1]{
\begin{tcolorbox}[colback=white,boxrule=1pt,top=0pt,bottom=0pt,left=1pt,right=2pt,top=2pt,bottom=2pt]
\em #1
\end{tcolorbox}
}

\AtBeginDocument{%
  \providecommand\BibTeX{{%
    \normalfont B\kern-0.5em{\scshape i\kern-0.25em b}\kern-0.8em\TeX}}}

\begin{document}

\title{\toolname: Code Translation with Corrector via LLMs}

\author{Xin Yin}
\affiliation{%
  \institution{Zhejiang University}
  \city{Hangzhou}
  \country{China}
  }
\email{xyin@zju.edu.cn}

\author{Chao Ni}
\authornote{This is the corresponding author}
\affiliation{
  \institution{Zhejiang University, Hangzhou High-Tech Zone (Binjiang) Blockchain and Data Security Research Institute}
  \city{Hangzhou}
  \country{China}
  }
\email{chaoni@zju.edu.cn}

\author{Tien N. Nguyen}
\affiliation{
  \institution{University of Texas at Dallas}
  \city{Texas}
  \country{USA}
  }
\email{tien.n.nguyen@utdallas.edu}

\author{Shaohua Wang}
\affiliation{
  \institution{Central University of Finance and Economics}
  \city{Bejing}
  \country{China}
  }
\email{davidshwang@ieee.org}

\author{Xiaohu Yang}
\affiliation{%
  \institution{Zhejiang University}
  \city{Hangzhou}
  \country{China}
  }
\email{yangxh@zju.edu.cn}

\begin{abstract}
Software migration is garnering increasing attention with the evolution of software and society. 
Early studies mainly relied on handcrafted translation rules to translate between two languages, the translation process is error-prone and time-consuming.
In recent years, researchers have begun to explore the use of pre-trained large language models (LLMs) in code translation.
However, code translation is a complex task that LLMs would generate mistakes during code translation, they all produce certain types of errors when performing code translation tasks, which include (1) compilation error, (2) runtime error, (3) functional error, and (4) non-terminating execution.
We found that the root causes of these errors are very similar (e.g. failure to import packages, errors in loop boundaries, operator errors, and more).
In this paper, we propose a general corrector, namely \toolname, which is a micro and universal model for repairing translation errors.
It learns from errors generated by existing LLMs and can be widely applied to correct errors generated by any LLM.
The experimental results on translation tasks between C++, Java, and Python show that our model has effective repair ability, and cross experiments also demonstrate the robustness of our method.
\end{abstract}



\maketitle

\section{Introduction}

{
Code translation is an important problem in software engineering. 
Translating code from one programming language to another enables reusing and porting software artifacts across languages and platforms. 
Early studies mainly relied on handcrafted translation rules to translate between two languages~\cite{2to3, C2Go, C2Rust}.
The translation is poor in readability and correctness, and needs extra manual corrections. 
Therefore, the translation process is error-prone and time-consuming~\cite{zhong2010mining}.
}

{
With the development of deep learning technologies, in
recent years, techniques based on Neural Machine Translation (NMT) have been extensively studied in recent years~\cite{chen2018tree, gu2017deepam, roziere2020unsupervised}.
These approaches treat translating code as an NMT problem, where the goal is to translate source code into target code and rely heavily on parallel training datasets obtained from open-source repositories.
However, parallel resources are much more scarce in the programming language domain than in natural language. 
It is costly to collect bilingual program data manually. Therefore, applying the NMT technology to code translation still faces many challenges.
}

{
To overcome the limitations of NMT-based approaches, researchers are exploring the use of pre-trained large language models (LLMs) for code translation, such as 
Codex~\cite{chen2021evaluating},
StarCoder~\cite{li2023starcoder},
CodeGen~\cite{nijkamp2022codegen}, CodeLlama~\cite{roziere2023code} and ChatGPT~\cite{openai2022chatgpt}, which generate correct code directly based on context by pre-training on large amounts of open-source code snippets.
Although prior works~\cite{pan2024lost, pan2023stelocoder} have shown promise in using LLMs for code translation, there is a dearth of research on understanding their limitations for this task. 
This is an important undertaking because code translation is a complex task that requires LLMs to understand code syntax (to generate syntactically correct code) and semantics (to preserve functionality during translation) simultaneously.
However, LLMs would produce certain types of errors when performing code translation tasks, which include (1) compilation error, (2) runtime error, (3) functional error, and (4) non-terminating execution. 
We found that the root causes of these errors are very similar (e.g. failure to import packages, errors in loop boundaries, operator errors, and more).
}

{
In this study, our objective is to enhance code translation through the introduction of a micro model exhibiting proficient error correction capabilities. 
This model can be applied universally to rectify errors arising from any LLM. 
To achieve this goal, we present \textbf{\toolname} with the following principal contributions.
Initially, we present a micro-level automated model tailored for rectifying translation errors. 
In contrast to LLMs, which demand substantial computational resources and associated costs, our micro model, fine-tuned on CodeT5+ 220M, necessitates significantly fewer resources than larger-scale LLMs such as Llama-2 13B.
Subsequently, we devise a universal model for rectifying errors produced by any LLM. 
Our model possesses a universal character, as it is not tailored to rectify errors specific to a particular LLM, but rather targets errors commonly encountered across different LLMs. 
This design is LLM agnostic and operates independently of any specific LLM architecture.
}

We conducted experiments on two extensively researched datasets, namely CodeNet~\cite{puri2021codenet} and AVATAR~\cite{liu2019avatar}, covering three highly prevalent programming languages: C++, Java, and Python.
These experiments involved a comparative assessment of four cutting-edge LLMs: ChatGPT~\cite{openai2022chatgpt}, StarCoder~\cite{li2023starcoder}, CodeGen~\cite{nijkamp2022codegen}, and CodeLlama~\cite{roziere2023code}.

Initially, we executed code translation tasks across all LLMs, with the outcomes strongly favoring ChatGPT. Specifically, on the CodeNet dataset, ChatGPT achieved an impressive success rate ranging from 59.5\% to 85.5\%. Likewise, on the AVATAR dataset, ChatGPT demonstrated the highest success rate, registering between 38.0\% and 73.1\%, which was notably 11.6\% to 61.8\% superior to its LLM counterparts.
Furthermore, the examined LLMs exhibited consistent patterns of translation errors, primarily manifesting as invalid code. These errors were manually rectified, resulting in valid code instances. These paired sets of valid and invalid codes were subsequently employed to fine-tune the CodeT5+ model. The results demonstrated the effectiveness of the CodeT5+ model fine-tuned on errors originating from ChatGPT, StarCoder, and CodeGen, effectively rectifying a total of 6 to 22 errors produced by CodeLlama.

Additionally, we conducted cross-experiments wherein we selected error code translations from ChatGPT, StarCoder, CodeGen, and CodeLlama sequentially, utilizing the error codes from the remaining models for fine-tuning. The experimental findings highlighted CodeT5+'s capacity to ameliorate translation errors across LLMs (i.e., 4.6\%$\sim$43.2\% of ChatGPT, 3.8\%$\sim$28.4\% of StarCoder, 3.9\%$\sim$21.3\% of CodeGen, and 6.4\%$\sim$24.4\% of CodeLlama).
This underscores the presence of similar error patterns across LLMs and affirms the universal, LLM-agnostic nature of \toolname, which operates independently of any specific LLM architecture.

In brief, the key contributions of this paper include:

\textbf{A. Comprehensive Evaluation of LLM-based Code Translation:} We perform a large-scale evaluation of the code translation using multiple LLMs. 
We consider the most recently released LLMs and our evaluation includes two crafted benchmarks spanning across C++, Java, and Python.

\textbf{B. \toolname: Micro and Universal Model for Repairing Translation Errors:}
We found that these LLMs generate similar error patterns during translation, so we manually corrected the errors generated by the LLMs and used a micro model to capture these error patterns.
The model fine-tuned on error data can be universally applied to any unknown LLMs.

\textbf{C. Extensive Empirical Evaluation:} 
We conducted cross experiments on 4 state-of-the-art LLMs on the widely studied CodeNet~\cite{puri2021codenet} and AVATAR~\cite{ahmad2021avatar} datasets to explore the effectiveness and robustness of \toolname.
The replication of this paper is publicly available~\cite{replication}.
\label{sec:intro}

\section{Motivation Example}

\subsection{A Motivation Example}

\begin{figure}[htbp]
    \centering
    \includegraphics[width=\linewidth]{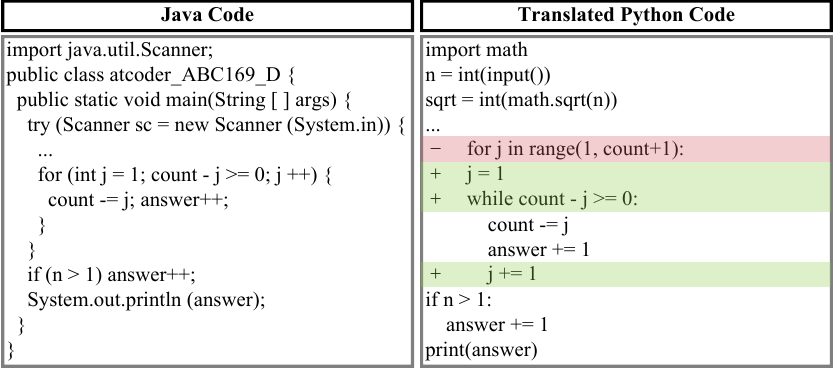}
    \caption{Translate the Java code ``atcoder\_ABC169\_D'' in the AVATAR dataset into Python code}
    \label{fig:motivation}
\end{figure}

Fig.~\ref{fig:motivation} shows the translation of the Java code ``atcoder\_ABC169\_D'' in the AVATAR dataset into Python code. 
The left part represents the Java code to be translated, while the right part represents the translation results of LLMs (i.e. ChatGPT, StarCoder, CodeGen, and CodeLlama).
This code is a solution to a programming problem on the AtCoder website.
The problem is given a positive integer N, consider repeatedly applying the operation below on N. 
First, choose a positive integer z satisfying all of the conditions below:
(1) z can be represented as z=$p^e$, where p is a prime number and e is a positive integer; 
(2) z divides N;
(3) z is different from all integers chosen in previous operations.
Then, replace N with N/z.
The solution code uses the $Scanner$ class to read input from the standard input stream and outputs the answer using the $System.out.println()$ statement.
These LLMs successfully translated the functionality of the original Java code, but they mistakenly translated the line $for\ (int\ j = 1;\ count\ -\ j \geq 0;\ j++)$
to $for\ j\ in\ range(1,\ count+1)$, which would result in different loop counts and incorrect results.
The correct translation should start from 1 and increase by 1 at each loop until $count\ -\ j$ equals 0.

\textbf{Observation 1.} 
\ul{LLMs would generate similar erroneous patterns during code translation.}
Over the years, several state-of-the-art LLMs~\cite{openai2022chatgpt, li2023starcoder, nijkamp2022codegen, wang2023codet5+, touvron2023llama, roziere2023code, chen2021evaluating, zheng2023codegeex} have been proposed, and show strong translation capabilities through pre-trained using millions of code snippets from open-source projects.
However, they all produce certain types of errors when performing code translation tasks, which include (1) compilation error, (2) runtime error, (3) functional error, and (4) non-terminating execution.
We found that the root causes of these errors are very similar (e.g. failure to import packages, errors in loop boundaries, operator errors, and more). 
By identifying common error types that repeatedly appear in the code, we can use unified correction operations to fix these errors, making the process of error correction more automated and reliable.

\textbf{Observation 2.} 
\ul{The existing neural machine translation based (NMT-based) models do not have the ability to generally correct errors, while using LLM to correct mistakes is relatively costly.}
Several SOTA NMT-based models~\cite{jiang2023knod, meng2023tenure, ye2022selfapr, ye2022neural, jiang2021cure} show strong error fixing capabilities through training on large amounts of labeled data. 
However, none of them has possessed powerful analytical reasoning capabilities to auto-fix the error shown in Fig.~\ref{fig:motivation}.
If there are no similar repair patterns in their limited training data, it becomes difficult to correctly fix the error, as none of them can understand and reason to add new logic into the code for fixing.
Unlike current NMT-based models using limited training data, the LLMs are directly pre-trained using millions of code snippets from open-source projects.
By utilizing high-quality prompts or fine-tuning, it can comprehend translation errors in the code and execute repairs~\cite{xia2023automated,xia2023keep}.
However, using LLM to correct translation errors requires significant computational resources at a high cost.

\subsection{Key Ideas}

Based on the above observations, we design our code translation framework with an automated corrector via LLMs, namely {\bf {\toolname}},  with the following key ideas.

(1) \textbf{Compact Error Correction Model.} 
We present an efficient error correction model capable of assimilating rectification patterns gleaned from translation errors generated by LLMs. This model exhibits the capacity to automatically rectify analogous errors caused by diverse LLMs. In contrast to LLMs, which entail substantial computational resources and associated costs, our compact model, fine-tuned on CodeT5+ 220M, demands significantly fewer resources than its LLM counterparts (e.g., Llama 2 13B).

(2) \textbf{Universal Model for Translation with Corrector.} 
When a novel LLM undertakes code translation, it tends to manifest comparable patterns of translation errors. Our model assimilates these patterns and can be applied to rectify errors generated by a spectrum of LLMs. In essence, our model boasts a universal applicability, as it is not tailored to rectify errors in any specific LLM but rather addresses common error patterns exhibited across various LLMs. This underscores its LLM-agnostic design paradigm.

\label{sec:motivate}

\section{\toolname: Code Translation with Corrector via LLMs}

\subsection{Collection Phase}

The purpose of this phase is to gather erroneous translations from the output of LLM.
These errors will serve as the foundation for identifying patterns of mistakes for the subsequent phase.
To achieve this, we need to address three tasks: (1) Prompt Preparation, (2) Translation Collection, and (3) Mistake Correction.

\subsubsection{Task 1: Prompt Preparation}

\begin{figure}[htbp]
    \centering
    \includegraphics[width=\linewidth]{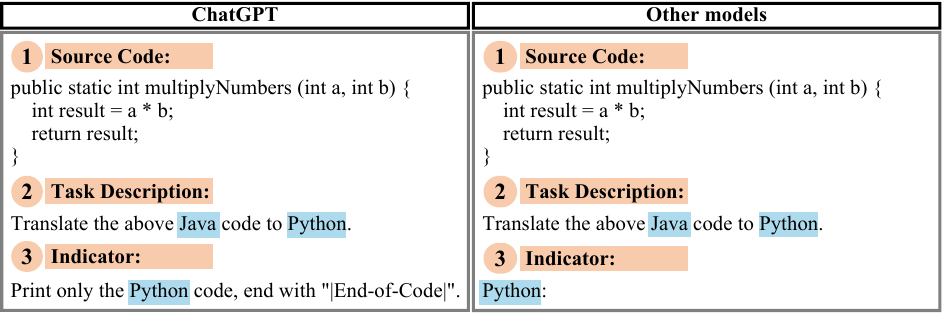}
    \caption{Prompt for ChatGPT and other models}
    \label{fig:prompt}
\end{figure}

We followed the prompt similar to those we found in the artifacts, papers, or technical reports associated with each model~\cite{nijkamp2022codegen, li2023starcoder, roziere2023code}.
The prompt used for LLMs involves three important components as illustrated in Fig.~\ref{fig:prompt}:

\begin{itemize}[leftmargin=*]
\item \textbf{Source Code} (marked as \ding{172}). 
We provide LLMs with code to be translated from different languages (i.e. Java, C++, and Python) in code translation task.
\item \textbf{Task Description} (marked as \ding{173}). LLMs are provided with the description constructed as \code{``Translate the above \$SOURCE\_LANG code to \$TARGET\_LANG.''}. 
The task descriptions used in the translation task vary based on the source and target programming languages we employ.
\item \textbf{Indicator} (marked as \ding{174}). 
ChatGPT outputs a large amount of descriptive text during inference.
Therefore, we need a strict prompt template to keep ChatGPT focused on the translation code rather than the descriptive text.
In this paper, we follow the best practice in previous work~\cite{pan2023understanding} and adopt the same prompt named \code{``Print only the \$TARGET\_LANG code, end with "|End-of-Code|".''}.
Other models are instructed to generate \code{``\$TARGET\_LANG''}.
\end{itemize}

\subsubsection{Task 2: Translation Collection}

Through the interaction of the Language Model (LLM) with the embellished prompt, it adheres to the task description in order to produce a translated code corresponding to the provided source code. It is worth noting that these generated translations might incorporate extraneous dialogue and descriptive text. To isolate the essential code segments, we employ regular expressions. As a result, the execution of Task 2 yields a compilation of translated codes from all LLMs.

\subsubsection{Task 3: Mistake Correction}

We utilize the test cases in the dataset to verify the accuracy of the code generated by LLM. If the translated code successfully passes all test cases, it is deemed a valid translation; otherwise, it is marked as invalid. There are four categories of translation errors: (1) compilation errors, (2) runtime errors, (3) functional errors, and (4) non-terminating executions.

Subsequently, we apply minor corrections to the invalid code in order to ensure it passes all test cases, resulting in what we refer to as valid code. Many of the errors produced by LLM can be rectified through straightforward adjustments, such as adding packages, modifying operators, or adjusting boundary conditions. The distinguishing factor between valid and invalid code lies in the specific erroneous statement, which aids the model in learning from its mistakes. These pairs of valid and invalid codes are then employed for error pattern learning in the subsequent phase.

\subsection{Fine-Tune and Inference Phase}

\begin{figure}[htbp]
    \centering
    \includegraphics[width=.8\linewidth]{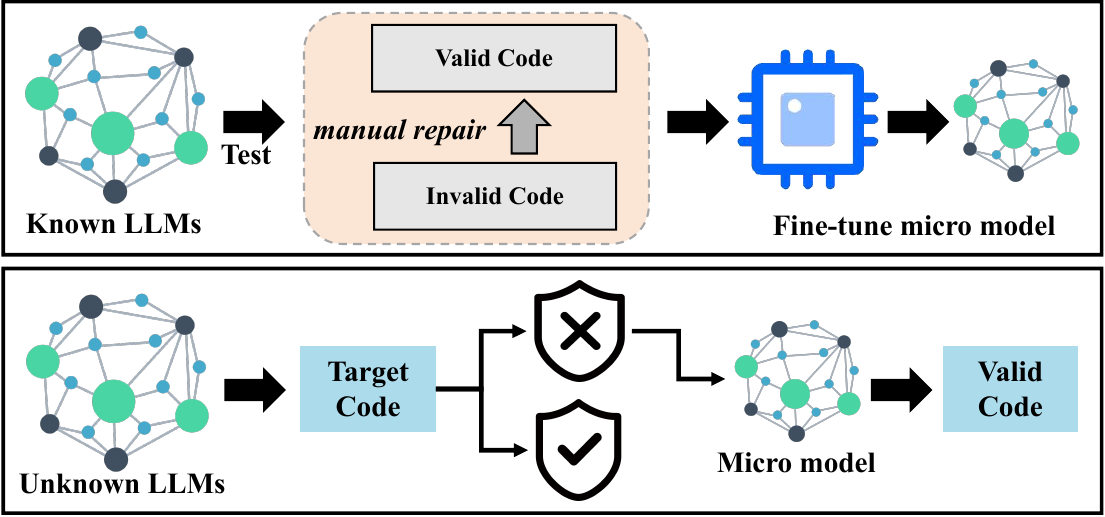}
    \caption{Fine-tune a smaller model to be the general corrector}
   \label{fig:inference}
\end{figure}

As illustrated in Fig.~\ref{fig:inference}, we employ the pairs of valid and invalid codes obtained during the collection phase to fine-tune a generated model. The purpose of fine-tuning is to assimilate the mistake patterns produced by established LLMs. The input to this generated model comprises the erroneous translation generated by LLMs, with the output being the corrected code.

For our study, we utilize the CodeT5+ model~\cite{wang2023codet5+} as the underlying LLM, although it can be readily substituted with various other generated LLMs, such as Llama 2~\cite{touvron2023llama}. Detailed fine-tuning procedures are elucidated in Section~\ref{sec:implementation}.

Following fine-tuning, CodeT5+ effectively learns the mistake patterns associated with LLMs in code translation, particularly within the training set. Subsequently, we utilize the fine-tuned CodeT5+ to correct the invalid code within the test set, which comprises models that CodeT5+ has not encountered previously (i.e., unknown LLMs, as depicted in Fig.~\ref{fig:inference}). Here, we input the erroneous translation generated by the unknown model into the fine-tuned CodeT5+, which then employs its learned correction pattern to find a solution for the current incorrect translation, ultimately producing the corrected code.

\label{sec:approach}

\section{Experimental Methodology}

\subsection{Dataset Collection and Pre-Processing}

In order to ensure the thoroughness and validity of our research findings regarding the nature of LLM translation errors, we have leveraged two widely recognized code translation benchmarks: the \textbf{CodeNet dataset~\cite{puri2021codenet}} and the \textbf{AVATAR dataset~\cite{ahmad2021avatar}}. These datasets have been previously employed in studies~\cite{pan2023understanding, szafraniec2022code} and cover three highly popular programming languages, namely C++, Java, and Python. The detailed characteristics of these selected datasets, along with their respective statistics, are shown in Table~\ref{tab:dataset}. Each of these datasets is equipped with test cases designed to validate code translations. Specifically, for CodeNet and AVATAR, the tests comprise input data and corresponding expected outputs.

\begin{table}[htbp]
  \centering
  \caption{Statistics of studied datasets}
  \resizebox{\linewidth}{!}
  {
    \begin{tabular}{lccccc}
    \toprule
    \textbf{Dataset} & \multicolumn{1}{l}{\textbf{Source Language}} & \textbf{\# Number} & \textbf{\# Testcase} & \textbf{Target Language} & \multicolumn{1}{l}{\textbf{\# Translation}} \\
    \multicolumn{1}{c}{\multirow{3}[1]{*}{CodeNet}} & C++ & 200 & 200 & Java, Python & 400 \\
      & Java & 200 & 200 & C++, Python & 400 \\
      & Python & 200 & 200 & C++, Java & 400 \\
    \midrule
    \multicolumn{1}{c}{\multirow{2}[2]{*}{AVATAR}} & Java & 249 & 6255 & C++, Python & 498 \\
      & Python & 250 & 6255 & C++, Java & 500 \\
    \midrule
    \textbf{\# Total} & \textbf{-} & \textbf{1099} & \textbf{13110} & \textbf{-} & \textbf{2198} \\
    \bottomrule
    \end{tabular}%
  }
  \label{tab:dataset}%
\end{table}%

\subsection{Studied Baseline Models}

\textit{\ul{Baselines.}} 
To comprehensively compare the performance of existing work, in this paper, we consider the four state-of-the-art LLMs, namely ChatGPT~\cite{openai2022chatgpt}, StarCoder~\cite{li2023starcoder}, CodeGen~\cite{nijkamp2022codegen}, and CodeLlama~\cite{roziere2023code}, and for error pattern corrector, we choose CodeT5+~\cite{wang2023codet5+} for our \toolname.
Here, we briefly introduce these methods to make our paper self-contained.

\textbf{ChatGPT} proposed by OpenAI~\cite{openai2022chatgpt} is a large pre-trained language model and is fine-tuned with the Reinforcement Learning with Human Feedback (RLHF) approach. It conducts multi-turn natural dialogs, comprehending history and generating coherent responses.
ChatGPT represents advanced language modeling and conversational AI.
It's key strengths include common sense reasoning and dialog coherence.


%
\textbf{StarCoder} proposed by Li et al.~\cite{li2023starcoder} is a large pre-trained language model specifically designed for code. 
It was pre-trained on a large amount of code data to acquire programming knowledge and trained on permissive data from GitHub, including over 80 programming languages, Git commits, GitHub issues, and Jupyter notebooks. 
StarCoder can perform code editing tasks, understand natural language prompts, and generate code that conforms to APIs. 
StarCoder represents the advancement of applying large language models in programming.

{\textbf{CodeGen} proposed by Nijkamp et al.~\cite{nijkamp2022codegen} is an AI system for generating code from natural language.
It utilizes a large pre-trained language model fine-tuned on programming data. 
CodeGen can translate natural language descriptions into working code in multiple languages.
CodeGen can be used to synthesize code that matches the specified functionality and integrate the generated code into the project.

{\textbf{CodeLlama}} proposed by Rozière et al.~\cite{roziere2023code} is a set of large pre-trained language models for code built on Llama 2.
They achieve state-of-the-art performance among open models on code tasks, provide infilling capabilities, support large input contexts, and demonstrate zero-shot instruction following for programming problems. 
CodeLlama is created by further training Llama 2 using increased sampling of code data. 
As with Llama 2, the authors applied extensive safety mitigations to the fine-tuned CodeLlama versions.

\textbf{CodeT5+} proposed by Wang et al.~\cite{wang2023codet5+} is a family of encoder-decoder models for code.
Its component modules can be combined in diverse ways to fit various downstream code tasks.
This flexibility comes from the mix of pre-training objectives designed by the authors to reduce the gap between pre-training and fine-tuning. 
These objectives include single-modal and dual-modal cross-lingual model pre-training tasks for cross-lingual code and text, such as span denoising, contrastive learning, and text-code matching.

\subsection{Experimental Procedure}
\label{sec:implementation}

{\textit{\ul{Data Splitting.}} 
We divided LLMs into two groups (i.e. LLM used for the error pattern corrector and LLM used for the code translator). 
For the LLM used for collecting errors, we adopt the data splitting approach: 80\%:10\%:10\%.
More precisely, the whole dataset is split into 80\% of training data, 10\% of validation data, and 10\% of testing data.
For the LLM used for inference, we take all the errors generated by the LLM as the testing data.
}

\textit{\ul{Model Implementation.}}
Regarding StarCoder, CodeGen, and CodeLlama, we utilize their publicly available source code and perform inference with the default parameters provided in their original code. 
All these models are implemented using the PyTorch~\cite{pytorch} framework by fully adopting the pre-trained models hosted on Huggingface~\cite{huggingface}. The fine-tuning process is performed on NVIDIA RTX 3090 graphics card.
Considering ChatGPT's code is not publicly available, we implemented translation in Python by wrapping the ChatGPT ability through its API support~\cite{2023chatgptendpoint} and adhere to the best-practice guide~\cite{shieh2023best} for each prompt.
We utilize the GPT-3.5-Turbo-0301 model from the ChatGPT family, which is the version used uniformly for our experiments.
\label{sec:experiment}

\section{Empirical Results}

To investigate the error patterns of different LLMs in code translation and evaluate our code translation framework with a corrector, our experiment focuses on the following research questions:

\begin{itemize}[leftmargin=*]
\item \textbf{RQ-1 Effectiveness of LLMs in Code Translation}. {\em How do state-of-the-art code LLMs perform in code translation across various benchmarks?}
\item \textbf{RQ-2 Category of Translation Errors}. {\em What are the different types of erroneous patterns for unsuccessful translation?}
\item \textbf{RQ-3 Effectiveness of \toolname in Error Repairing}. 
{\em (1) Can the patterns learned from existing errors be used to fix errors generated by unknown LLM?}
{\em (2) How do different sources of errors affect the overall performance of the model (i.e., the robustness of model)?} 
\end{itemize}

\subsection{RQ-1: Effectiveness of LLMs in Code Translation}
\label{sec:rq1}

\textbf{\underline{RQ1-Analysis Procedure.}}
In this research, we delineate four categories of translation errors: (1) compilation errors, (2) runtime errors, (3) functional errors, and (4) instances of non-terminating execution. We deliberately exclude static evaluation metrics such as exact match, syntax match, dataflow match~\cite{ren2020codebleu}, CodeBLEU~\cite{ren2020codebleu}, and CrystalBLEU~\cite{eghbali2022crystalbleu}, as our primary objective is to verify the viability of the translations through compilation and execution. It is worth noting that static metrics can potentially be misleading in the context of code synthesis~\cite{chen2021evaluating}. Specifically, language models may yield seemingly favorable scores on these metrics, yet produce code that proves inexecutable due to compilation or runtime issues~\cite{ahmad2021avatar, chen2021evaluating}.

\textbf{\underline{RQ1-Results.}} {\bf Performance of LLMs in translating code.} 
Table~\ref{tab:performance} shows the detailed results of LLMs for code translation. 
We can observe that: (1) ChatGPT, StarCoder, and CodeLlama perform far better than CodeGen, especially ChatGPT achieving the best performance (except for Java $\rightarrow$ Python translation on the CodeNet dataset), with translation success rates 38.0\%$\sim$85.5\%.
(2) When LLM performs translation C++ $\rightarrow$ Java or Java $\rightarrow$ C++, it usually achieves better translation effects, such as translation C++ $\rightarrow$ Java on the CodeNet dataset (i.e., 85.0\%), translation Java $\rightarrow$ C++ on the CodeNet dataset (i.e., 85.5\%), and translation Java $\rightarrow$ C++ on the AVATAR dataset (i.e., 73.1\%), which indicates that LLM is better at translating for the languages of the same type (e.g., Java and C++ are both static languages). 
(3) The translation performance of LLM on the AVATAR dataset is lower than that on the CodeNet dataset, which is due to a strong correlation between the translation success rate and the number of test cases in the dataset (i.e., 200 test cases in CodeNet dataset and 6,255 test cases in AVATAR dataset, shown in Table~\ref{tab:dataset}). That is, the more stricter the existing test suite is, the better the evaluation of whether the translation has successfully preserved the functionality.

\begin{table}[htbp]
  \centering
  \caption{Performance of LLMs in translating code from different studied datasets}
  \resizebox{\linewidth}{!}
  {
    \begin{tabular}{lccccccc}
    \toprule
    \multirow{2}[1]{*}{\textbf{Dataset}} & \multirow{2}[1]{*}{\textbf{Source Language}} & \multirow{2}[1]{*}{\textbf{Target Language}} & \multirow{2}[1]{*}{\textbf{\# Number}} & \multicolumn{4}{c}{\textbf{\% Successful Translation}} \\
          &       &       &       & \multicolumn{1}{c}{\textbf{ChatGPT}} & \multicolumn{1}{c}{\textbf{StarCoder}} & \multicolumn{1}{c}{\textbf{CodeGen}} & \multicolumn{1}{c}{\textbf{CodeLlama}} \\
    \midrule
    \multirow{6}[2]{*}{\textbf{CodeNet}} & C++ & Java & 200 & \cellcolor{lightgray}\textbf{85.0\%} & 63.5\% & 3.5\% & 67.0\% \\
          & C++ & Python & 200   & \cellcolor{lightgray}\textbf{62.0\%} & 33.0\% & 6.0\% & 36.0\% \\
          & Java  & C++   & 200   & \cellcolor{lightgray}\textbf{85.5\%} & 60.0\% & 32.0\% & 65.5\% \\
          & Java  & Python & 200   & \cellcolor{lightgray}\textbf{57.5\%} & 25.0\% & 6.0\% & 43.0\% \\
          & Python & C++   & 200   & \cellcolor{lightgray}\textbf{80.5\%} & 57.0\% & 35.5\% & 62.5\% \\
          & Python & Java  & 200   & 59.5\% & \cellcolor{lightgray}\textbf{61.0\%} & 0.5\% & 55.0\% \\
    \midrule
    \multirow{4}[2]{*}{\textbf{AVATAR}} & Java  & C++   & 249   & \cellcolor{lightgray}\textbf{73.1\%} & 35.7\% & 12.4\% & 47.4\% \\
          & Java  & Python & 249   & \cellcolor{lightgray}\textbf{62.2\%} & 16.1\% & 0.4\% & 31.7\% \\
          & Python & C++   & 250   & \cellcolor{lightgray}\textbf{38.8\%} & 26.8\% & 7.2\% & 24.8\% \\
          & Python & Java  & 250   & \cellcolor{lightgray}\textbf{38.0\%} & 26.4\% & 1.6\% & 24.8\% \\
    \bottomrule
    \end{tabular}%
  }
  \label{tab:performance}%
\end{table}%

{\bf Breakdown of Unsuccessful Translations.}
The prior findings indicate that a majority of Large Language Models (LLMs) demonstrate satisfactory performance in the realm of code translation when assessed on meticulously designed benchmarks. Toward our goal, we subsequently categorize unsuccessful translations based on their respective error outcomes, which encompass: (1) Compilation Error, denoting instances where the translated code cannot be successfully compiled; (2) Runtime Error, signifying scenarios in which the translated code compiles but subsequently encounters a runtime exception; (3) Functional Error, characterizing cases where the translated code compiles and executes without error, yet yields a test failure due to output discrepancies compared to the source program; and (4) Non-terminating Execution, referring to situations in which the translated code successfully compiles and initiates execution, but fails to terminate, often due to an encounter with an infinite loop or a waiting state for user input.

Table~\ref{tab:outcomes} and Fig.~\ref{fig:outcomes} show the breakdown of the unsuccessful translations produced by LLMs for each dataset and the proportion of translation results for each LLM.
We observe that: (1) The proportion of compilation errors generated by translation is the highest (i.e. 36.9\%$\sim$68.2\% shown in Table~\ref{tab:outcomes} and 42.3\%$\sim$60.2\% shown in Fig.~\ref{fig:outcomes}), which indicated that these LLMs are difficult to understand the target code syntax. 
(2) Further breakdown of the results per PLs shows that Java and C++ have comparatively stricter syntax, while it is easier for LLMs to generate syntactically correct Python code.
(3) The next most common effect of unsuccessful translation
is a functional error (i.e., 12.2\%$\sim$46.7\% shown in Table~\ref{tab:outcomes} and 21.6\%$\sim$31.2\% shown in Fig.~\ref{fig:outcomes}), demonstrating that even when the code is syntactically correct and terminates with no exception or runtime error, it does not maintain the functionality implemented in the source language.


\begin{table}[htbp]
  \centering
  \caption{Breakdown of the unsuccessful translations produced by LLMs for each dataset}
  \resizebox{\linewidth}{!}
  {
    \begin{tabular}{l|rr|rr|rr|rr|rr}
    \toprule
    Dataset & \multicolumn{6}{c|}{CodeNet}                  & \multicolumn{4}{c}{AVATAR} \\
    \midrule
    Source Language & \multicolumn{2}{c|}{C++} & \multicolumn{2}{c|}{Java} & \multicolumn{2}{c|}{Python} & \multicolumn{2}{c|}{Java} & \multicolumn{2}{c}{Python} \\
    Target Language & \multicolumn{1}{c}{Java} & \multicolumn{1}{c|}{Python} & \multicolumn{1}{c}{C++} & \multicolumn{1}{c|}{Python} & \multicolumn{1}{c}{C++} & \multicolumn{1}{c|}{Java} & \multicolumn{1}{c}{C++} & \multicolumn{1}{c|}{Python} & \multicolumn{1}{c}{C++} & \multicolumn{1}{c}{Java} \\
    \midrule
    Compilation Error & \cellcolor{lightgray}\textbf{68.2\%} & 47.5\% & 66.5\% & 39.1\% & 61.0\% & 64.7\% & 55.7\% & 36.9\% & 48.9\% & 50.1\% \\
    Runtime Error & 19.1\% & 33.7\% & 1.3\% & \cellcolor{lightgray}\textbf{46.6\%} & 0.6\% & 22.8\% & 1.6\% & 38.4\% & 1.3\% & 25.4\% \\
    Functional Error & 12.2\% & 18.4\% & 31.0\% & 13.8\% & 37.2\% & 12.1\% & 40.1\% & 23.7\% & \cellcolor{lightgray}\textbf{46.7}\% & 23.4\% \\
    Non-terminating Execution & 0.6\% & 0.4\% & 1.3\% & 0.6\% & 1.2\% & 0.4\% & 2.6\% & 1.0\% & \cellcolor{lightgray}\textbf{3.0\%} & 1.2\% \\
    \bottomrule
    \end{tabular}%
  }
  \label{tab:outcomes}%
\end{table}%

\begin{figure}[htbp]
    \centering
    \includegraphics[width=\linewidth]{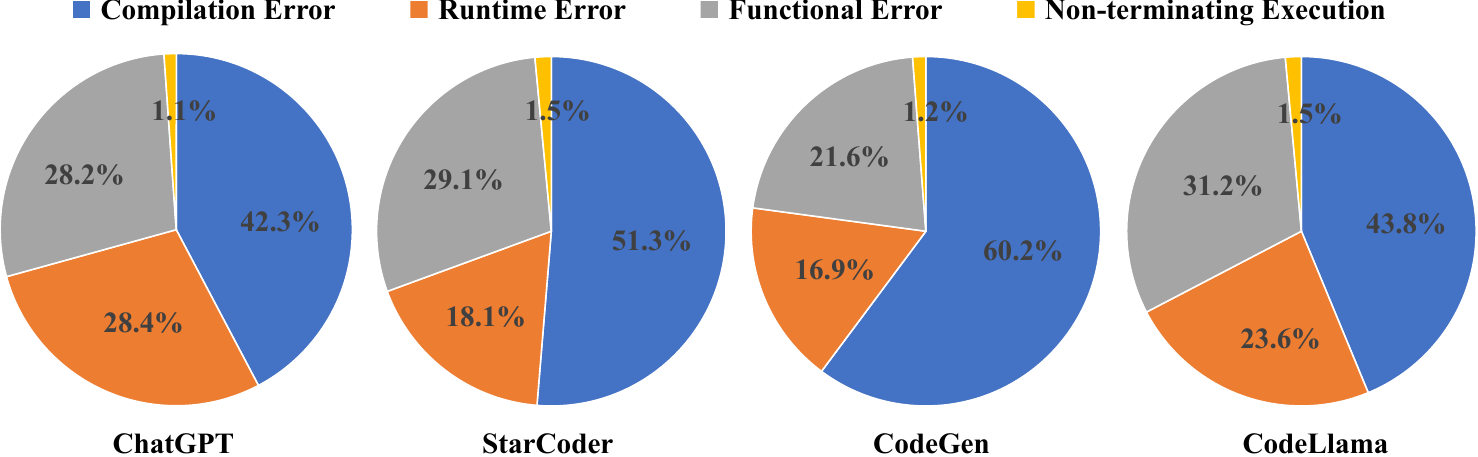}
    \caption{Proportion of translation results for each LLM}
    \label{fig:outcomes}
\end{figure}

\intuition{
{\bf Answer to RQ-1}: ChatGPT, StarCoder, CodeGen, CodeLlama perform code translation with several types of translation errors ranging from compiling, runtime, functional, and non-terminating errors in different degrees.
}

\subsection{RQ-2: Category of Translation Errors}
\label{sec:rq2}

\textbf{\underline{RQ2-Analysis Procedure.}}
To gain insights into the nature of translation anomalies, a rigorous investigation was conducted involving a manual scrutiny of the fundamental causes underlying unsuccessful translations. This inquiry is structured around the three above research questions, culminating in the establishment of an inclusive classification system for translation errors. Additionally, the study probes into the prevalence and spatial dispersion of each error category within the domain of unsuccessful translations. To streamline the manual labor involved in error comprehension and categorization, our attention was directed towards 5,342 instances of unsuccessful translations emanating from ChatGPT, StarCoder, CodeGen, and CodeLlama. The construction of the error classification system engaged the collaborative efforts of four human annotators, each with expertise in research or software engineering. 
The four annotators (not the authors) examine 5,342 errors of generated code. For each error, the four annotators independently study the translation error and classify it. When the labeling is finished, the annotators then compare their results and discuss each disagreement until reaching a consensus. We have a Cohen's Kappa~\cite{cantor1996sample} value of 0.80 in this process, which indicates a substantial agreement. This endeavor encompassed unsuccessful translations across all ten translation pairs outlined in Table~\ref{tab:dataset}.

\textbf{\underline{RQ2-Results.}}
We produced a category organized into six groups of root causes (Table~\ref{tab:category}): (1) Syntactic difference between languages, (2) Semantic difference between languages, (3) Dependency error, (4) Logic error, (5) Data-related error, (6) Model-specific error, and (7) Others.
In the rest of this section, we discuss the category groups with illustrative examples.

\begin{table}[htbp]
  \centering
  \caption{Categories of errors introduced during code translation by LLM}
  \resizebox{\linewidth}{!}
  {
    \begin{tabular}{l|cccc}
    \toprule
    \textbf{Category of Translation Errors} & \textbf{ChatGPT} & \textbf{StarCoder} & \textbf{CodeGen} & \textbf{CodeLlama} \\
    \midrule
    \textbf{Syntactic difference between languages} & 24.4\% & 29.0\% & \cellcolor{lightgray}\textbf{30.1\%} & 26.5\% \\
    \textbf{Semantic difference between languages} & 1.2\% & 1.0\% & \cellcolor{lightgray}\textbf{1.7\%} & 1.3\% \\
    \textbf{Dependency error} & \cellcolor{lightgray}\textbf{16.8\%} & 10.3\% & 15.2\% & 14.7\% \\
    \textbf{Logic error} & 8.0\% & 9.8\% & 8.5\% & \cellcolor{lightgray}\textbf{11.8\%} \\
    \textbf{Data-related error} & \cellcolor{lightgray}\textbf{43.9\%} & 25.4\% & 23.3\% & 27.0\% \\
    \textbf{Model-specific error} & 0.7\% & \cellcolor{lightgray}\textbf{20.7\%} & 16.5\% & 14.2\% \\
    \textbf{Others} & \cellcolor{lightgray}\textbf{4.9\%} & 3.8\% & 4.7\% & 4.5\% \\
    \bottomrule
    \end{tabular}%
  }
  \label{tab:category}%
\end{table}%

\subsubsection{\textbf{Syntactic difference between languages}} In this group, a set of discrepancies is evident, primarily attributed to the inefficacy of Language LLMs in proficiently managing syntactic disparities among Programming Languages (PLs) in code translation. {\em LLMs frequently emulate the syntax of the source PL, even when it is incompatible with the target PL}.

\begin{figure}[htbp]
    \centering
    \includegraphics[width=.8\linewidth]{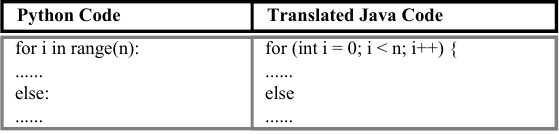}
    \caption{An example of the syntactic difference between languages}
    \label{fig:category_1}
\end{figure}

For instance, as illustrated in Fig.~\ref{fig:category_1}, an instance of an erroneous translation from Python to Java is depicted. In this instance, the LLM erroneously incorporates the \textit{for...else} loop from the source language, a construct not permissible within Java syntax.

\subsubsection{\textbf{Semantic difference between languages}}
Specifically, common errors include mismatches in API behaviors and incorrect use of operators. 
LLMs may incorrectly map source APIs to target PL methods, leading to code that does not properly execute.
Similarly, different PLs may have different operator syntax, leading to incorrect translations that can cause unexpected errors. 
As shown in Fig.~\ref{fig:category_2}, in the case where both the divisor and dividend are integers, / in Java represents integer division, while in Python it represents regular division.

\begin{figure}[htbp]
    \centering
    \includegraphics[width=.8\linewidth]{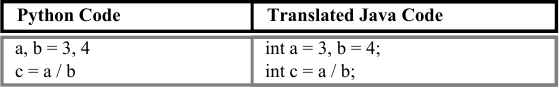}
    \caption{An example of semantic difference between languages}
    \label{fig:category_2}
\end{figure}

\subsubsection{\textbf{Dependency error}}
Import statements load necessary libraries, classes, and modules utilized in code. 
We found translation often leads to missing or incorrect imports, resulting in errors.
among many errors, LLMs struggle to translate definitions and implementations of data types, methods, etc. when imports are wrong.

\subsubsection{\textbf{Logic error}} 
When LLM performs code translation, it may misunderstand the logic of the source code and generate incorrect translation logic. 
This category covers: (1) incorrect loop and conditional boundaries, (2) inclusion of logic not in the source code, and (3) removal of logic in the source code.
Changes made to the logic of the source code will lead to differences in functionality.

\subsubsection{\textbf{Data-related error}} 
We observed numerous errors stemming from incorrect translation in data handling-including input parsing, data types, and output formatting. 
Specifically, LLMs failed to correctly parse and extract values from inputs, chose inappropriate data types for variables and return values, and formatted outputs incorrectly.

\subsubsection{\textbf{Model-specific error}}
Certain errors stem from inherent limitations in LLM design. 
For example, we have found some issues where the LLM does not output any target language code during code translation, outputs a large amount of duplicate code, or the token size of the LLM exceeds, resulting in compilation errors or no output being generated.

Table~\ref{tab:category} provides an exhaustive breakdown of translation errors. Our observations are as follows:

\begin{itemize}
    \item Predominantly, ChatGPT exhibits errors pertaining to data handling, constituting approximately 43.9\% of the total errors. These primarily manifest as input/output discrepancies. Fortunately, these errors are amenable to correction through pattern-based learning.
    \item Model-specific errors denote discrepancies unique to the Large Language Model (LLM), such as code output in a non-target language. These errors typically pose a greater challenge for resolution. Notably, ChatGPT exhibits a substantially lower incidence of such errors compared to other LLMs, underscoring its heightened resilience in code translation.
\end{itemize}

\intuition{
{\bf Answer to RQ-2}: The errors produced by an LLM in code translation follow different categories. They tend to have patterns that are amenable to correction through learning. 
}

\subsection{RQ-3: Effectiveness of Universal Model (\toolname) in Error Correction}
\label{sec:rq3}

\textbf{\underline{RQ3-Analysis Procedure.}}
Our results from RQ1 show that a majority of the translations by LLMs are unsuccessful due to the introduction of different errors, resulting in compilation errors, runtime errors, functional errors, and non-terminating execution.
In this section, we investigate whether our model learned from these errors can be used to repair translation errors generated by unknown (new) models.
We manually fixed the translation errors generated by the known LLMs (i.e., ChatGPT, StarCoder, and CodeGen), using a series of invalid-valid pairs to fine-tune the CodeT5+ model of \toolname.
Then, we evaluate whether the fine-tuned CodeT5+ 
learning from similar error patterns can repair the incorrect translation generated by an unknown model (i.e., CodeLlama).

To explore the universality of our proposed framework and its ability to run independently of any specific LLM model. 
We conducted cross experiment wherein we selected error code translations from ChatGPT, StarCoder, CodeGen, and CodeLlama sequentially, utilizing the error codes from the remaining models for fine-tuning.
Essentially, in this experiment, we conducted three additional experiments: (1) learn from StarCoder, CodeGen, and CodeLlama, test in ChatGPT, (2) learn from ChatGPT, CodeGen, and CodeLlama, test in StarCoder, and (3) learn from ChatGPT, StarCoder and CodeLlama, test in CodeGen.

\textbf{\underline{RQ3-Results.}} 
{Table~\ref{tab:repair} shows that our correction model of \toolname can repair translation errors generated by CodeLlama.} 
Our model learns from errors generated by ChatGPT, StarCoder, and CodeGen, and can fix errors generated by CodeLlama models that have not been learned before. 
Specifically, it can fix 22 out of 90 errors generated in Python$\rightarrow$Java translation in the CodeNet dataset.
Overall, we provide an effective translation-error-repairing model that can fix errors in the range of 6.4\% (12 out of 188) $\sim$24.4\% (22 out of 90).
The above results indicate that LLM generates similar errors in code translation tasks.
The patterns learned from existing code translation errors can be used to fix errors generated by new LLM.

\begin{table}[htbp]
  \centering
  \caption{Performance of repairing error translated code}
  \resizebox{\linewidth}{!}
  {
    \begin{tabular}{lcccrr}
    \toprule
    \textbf{Dataset} & \textbf{Source Language} & \textbf{Target Language} & \textbf{\# Number} & \textbf{\# Invalid} & \textbf{\# Repair} \\
    \midrule
    \multirow{6}[1]{*}{\textbf{CodeNet}} & C++   & Java  & 200   & 66    & 6 (9.1\%) \\
          & C++   & Python & 200   & 128   & 10 (7.8\%) \\
          & Java  & C++   & 200   & 69    & 8 (11.6\%) \\
          & Java  & Python & 200   & 114   & 16 (14.0\%) \\
          & Python & C++   & 200   & 75    & 7 (9.3\%) \\
          & Python & Java  & 200   & 90    & \cellcolor{lightgray}\textbf{22 (24.4\%)} \\
    \midrule
    \multirow{4}[2]{*}{\textbf{AVATAR}} & Java & C++   & 249 & 131 & 15 (11.5\%) \\
          & Java & Python & 249 & 170 & 13 (7.6\%) \\
          & Python & C++ & 250 & 188 & 12 (6.4\%) \\
          & Python & Java & 250 & 188 & 18 (9.6\%) \\
    \bottomrule
    \end{tabular}%
  }
  \label{tab:repair}%
\end{table}%

We also wanted to understand how translation errors evolve during this repairing process. 
To that end, we tracked the error outcomes of unsuccessful translations to further illustrate the effectiveness of error repair. For a better presentation, we used pie charts to represent the distribution of each error type after the repair operation. 
If an error is repaired and successfully passes all test cases, it is considered as a success.

Fig.~\ref{fig:changes} illustrates the results of our analysis of CodeLlama and we make the following observations:

\begin{enumerate}
    \item The model is more sensitive to compilation errors, and can successfully fix 13.5\% of this error. 
This is because LLM produces a large number of compilation errors during translation (cf. Section~\ref{sec:rq1}), so the majority of the compilation error samples in the fine-tuning dataset for CodeT5+ are included.

    \item For other errors, the successful percentage is lower (i.e., 7.1\% for Runtime Error and 7.1\% for Functional Error)-suggesting these errors are harder to mitigate.

    \item We also observe a few cases where the outcome of the translation upgrades (i.e., Compilation Error transforms to Runtime/Functional Error).
The model cannot fully restore the functionality of the code, but it still repairs some errors that currently exist in these codes.

\end{enumerate}

\begin{figure}[htbp]
    \centering
    \includegraphics[width=\linewidth]{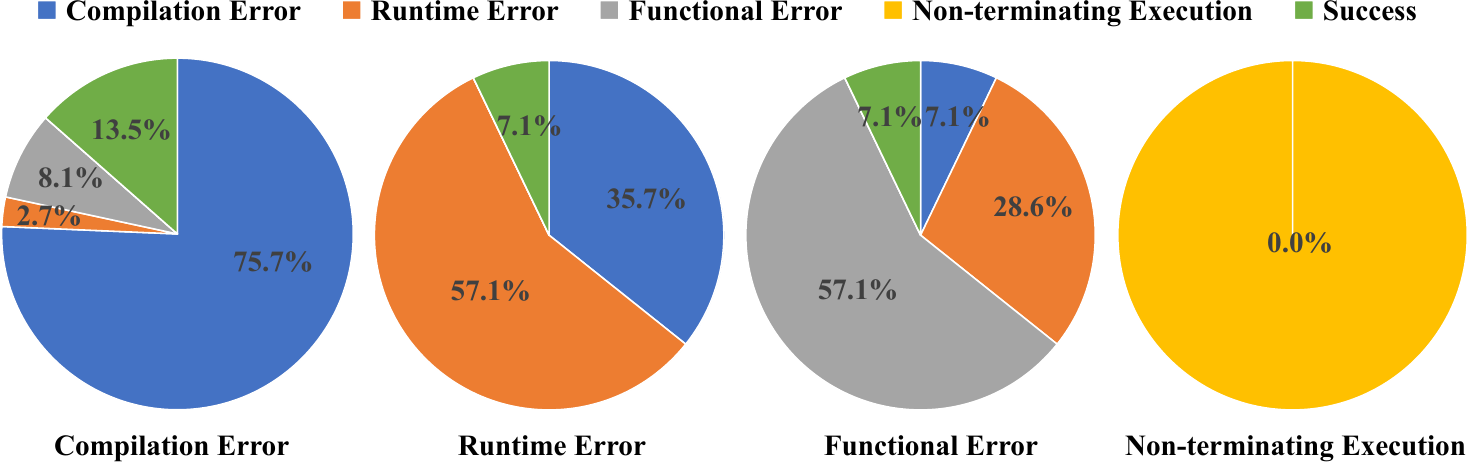}
    \caption{Error breakdown changes in Python$\rightarrow$Java translation on CodeNet dataset of CodeLlama}
    \label{fig:changes}
\end{figure}

\underline{\textbf{Cross experiment.}}
In order to investigate how different error sources affect the overall performance of the model (i.e. the robustness of the model), we conducted a cross experiment. According to the results in Table~\ref{tab:cross}, we can observe that:

\begin{enumerate}
    \item Our model is more sensitive to errors generated by ChatGPT.
In RQ2 (cf. Section~\ref{sec:rq2}), we mentioned that 43.9\% of errors generated by ChatGPT are due to data-related errors, particularly input/output format errors. 
These types of errors also frequently appear in other LLMs, therefore, through effective error pattern learning, our model can fix 4.6\%$\sim$43.2\% of ChatGPT errors.
  \item Although StarCoder, CodeGen, and ChatGPT generate similar error patterns, StarCoder and CodeGen produce a large number of model-specific errors when translating code, such as outputting code that is irrelevant to the target language. 
This type of error is difficult to repair, resulting in weaker repair performance of our model on StarCoder and CodeGen (i.e., 3.8\%$\sim$28.4\%).
\end{enumerate}

\begin{table}[htbp]
  \centering
  \caption{Performance of repairing each LLM in cross experiment}
  \resizebox{\linewidth}{!}
  {
    \begin{tabular}{lccrrr}
    \toprule
    \multicolumn{1}{r}{\multirow{2}[2]{*}{\textbf{Dataset}}} & \multirow{2}[2]{*}{\textbf{Source Language}} & \multirow{2}[2]{*}{\textbf{Target Language}} & \multicolumn{3}{c}{\textbf{\# Invalid / \# Repair}} \\
         \cmidrule{4-6} &       &       & \textbf{ChatGPT} & \textbf{StarCoder} & \textbf{CodeGen} \\
    \midrule
    \multicolumn{1}{r}{\multirow{6}[2]{*}{\textbf{CodeNet}}} & C++   & Java  & 30/3 (10.0\%) & \cellcolor{lightgray}\textbf{73/10 (13.7\%)} & 193/19 (9.8\%) \\
          & C++   & Python & 76/17 (22.4\%) & \cellcolor{lightgray}\textbf{134/38 (28.4\%)} & 188/40 (21.3\%) \\
          & Java  & C++   & \cellcolor{lightgray}\textbf{29/4 (13.8\%)} & 80/6 (7.5\%) & 136/10 (7.4\%) \\
          & Java  & Python & \cellcolor{lightgray}\textbf{85/28 (32.9\%)} & 150/38 (25.3\%) & 188/10 (5.3\%) \\
          & Python & C++   & \cellcolor{lightgray}\textbf{39/7 (17.9\%)} & 86/10 (11.6\%) & 129/9 (7.0\%) \\
          & Python & Java  & \cellcolor{lightgray}\textbf{81/35 (43.2\%)} & 78/8 (10.3\%) & 199/20 (10.1\%) \\
    \midrule
    \multicolumn{1}{r}{\multirow{4}[2]{*}{\textbf{AVATAR}}} & Java  & C++   & \cellcolor{lightgray}\textbf{67/6 (9.0\%)} & 160/9 (5.6\%) & 218/19 (8.7\%) \\
          & Java  & Python & \cellcolor{lightgray}\textbf{94/11 (11.7\%)} & 209/16 (7.7\%) & 248/20 (8.1\%) \\
          & Python & C++   & \cellcolor{lightgray}\textbf{153/7 (4.6\%)} & 183/7 (3.8\%) & 232/9 (3.9\%) \\
          & Python & Java  & \cellcolor{lightgray}\textbf{155/31 (20.0\%)} & 184/12 (6.5\%) & 246/17 (6.9\%) \\
    \bottomrule
    \end{tabular}%
  }
  \label{tab:cross}%
\end{table}%

\intuition{
{\bf Answer to RQ-3}: 
Our model demonstrates strong capability in repairing errors from unknown language models, successfully fixing translation errors generated by ChatGPT, CodeGen, StarCoder, and CodeLlama showing high robustness across diverse error patterns produced by different LLMs.
}
\label{sec:results}

\section{Case Studies}

To further understand why our \textbf{\toolname} performs well in correcting translation errors, we further analyzed some examples as case studies, including (1) syntactic differences between languages, (2) semantic differences between languages, (3) dependency error, and (4) data-related error.
We also present two examples of translation errors that are difficult to repair.
We will elaborate on the causes of each error and demonstrate how our proposed model repairs them in detail.

\subsection{Syntactic differences between languages}
We analyze the example of the error on syntactic differences between languages (i.e., atcoder\_ABC174
\_C.py in AVATAR) in Fig.~\ref{fig:example_1}.
The purpose of this Python code is to find the smallest integer $i$, when represented in base 10 with every digit replaced by 7, results in a number that is divisible by $k$.
The loop simulates the replacement and verification process. 
$t$ stores the current test value, and checks if it is divisible by $k$. 
If it is, then the number of iterations $i$ is the answer. 
If the loop ends and no feasible value of $i$ is found, then there is no such $i$, and -1 is printed.
Overall, this code intelligently searches for the smallest integer $i$ that satisfies a specific condition through simulating base conversion and modulo operations.
There are compilation error in the translated C++ code, the StarCoder erroneously incorporates the \textit{for...else} loop from the source language, a construct not permissible within C++ syntax. 
Our proposed model successfully identified this error, and the modified C++ code has exactly the same functionality as the Python code, passing all test cases.

\begin{figure}[htbp]
    \centering
    \includegraphics[width=\linewidth]{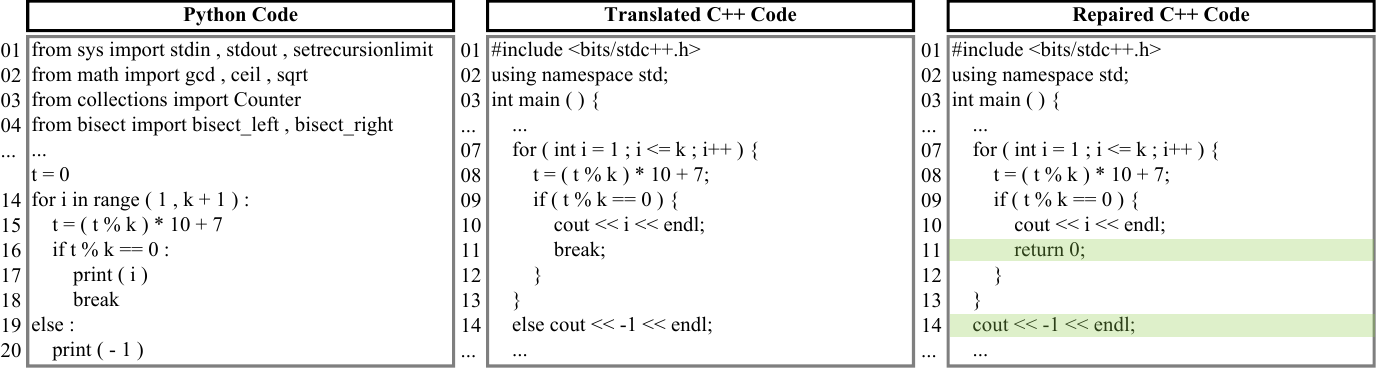}
    \caption{Syntactic differences between languages of StarCoder on AVATAR}
    \label{fig:example_1}
\end{figure}

\subsection{Semantic differences between languages}

We analyze the example of semantic differences between languages error (i.e., s369598583.Java in CodeNet) in Fig.~\ref{fig:example_2}.
The function of this Java code is to calculate the sum of the digits of all positive integers between 1 and N.
Specifically, it uses a scanner to obtain an integer N, then uses a loop to calculate the sum of the digits of all numbers between 1 and N, and finally outputs the result. 
In the calculation process, it uses a counter count and a variable next to update the range of the loop calculation according to a specific pattern.
CodeLlama translated this Java code into Python code. 
Although it did not produce any compilation errors, it did not realize the difference between for loops in Java and Python languages.
In Fig.~\ref{fig:example_5}, line 8 of the Java code is a $for$ loop, while line 6 of the translated Python code is a $for$ loop.
In Python's $for$ loop, $i$ does not act as a counter loop variable, but refers to the elements in the iterable object range.
Therefore, when $i \textgreater N$ in the Python code, the loop statement is still executed, which is different from the functionality of the Java code. 
Our proposed model successfully fixed this error by adding $if \ i \textgreater N:\ break$ statement in the for loop. 
Therefore, if $i \textgreater N$, the loop is exited.

\begin{figure}[htbp]
    \centering
    \includegraphics[width=\linewidth]{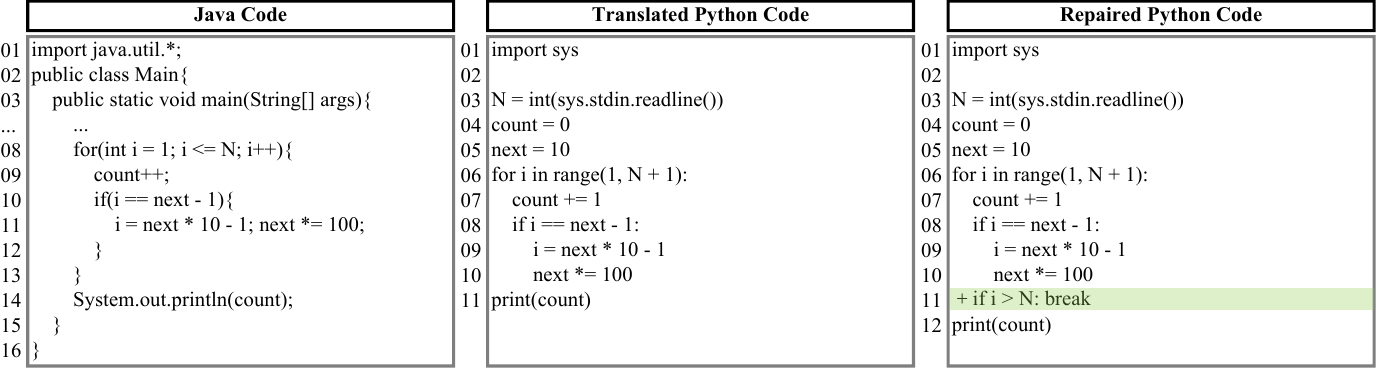}
    \caption{Semantic differences between languages of CodeLlama on CodeNet}
    \label{fig:example_2}
\end{figure}

\subsection{Dependency error}

We analyzed the example of dependency error (i.e., codeforces\_421\_A.py in AVATAR) in Fig.~\ref{fig:example_3}.
This error is a representative example of dependency errors, which occurs because Scanner class is used in Java code but the relevant package is not imported.
Our model adds the import statement to successfully repair the error.

\begin{figure}[htbp]
    \centering
    \includegraphics[width=\linewidth]{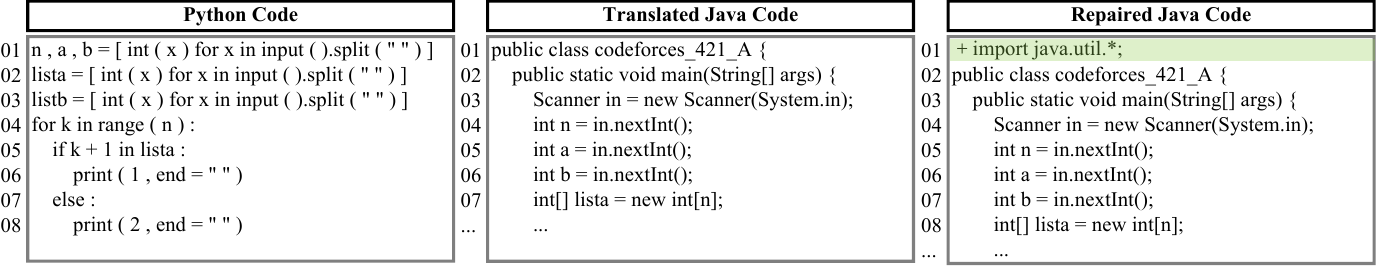}
    \caption{Dependency error of CodeGen on AVATAR}
    \label{fig:example_3}
\end{figure}

\subsection{Data-related error}

The investigation centered on a data-related anomaly (i.e., s987117545.java in CodeNet) illustrated in Fig.~\ref{fig:example_4}. This anomaly stemmed from an input parsing discrepancy within the translated Python code. Specifically, the test case's input comprised two data elements situated on a single line, delimited by a space. Utilizing the $input()$ function led to the concurrent retrieval of both data items from the same line. Thus, adopting $map(int, input().split())$ for the reading operation became imperative to capture both data items accurately. This category of anomaly was found to be prevalent across numerous translations produced by other LLMs. Consequently, a model attuned to identifying and rectifying such error patterns could proficiently address this particular type of anomaly generated by ChatGPT.

\begin{figure}[htbp]
    \centering
    \includegraphics[width=\linewidth]{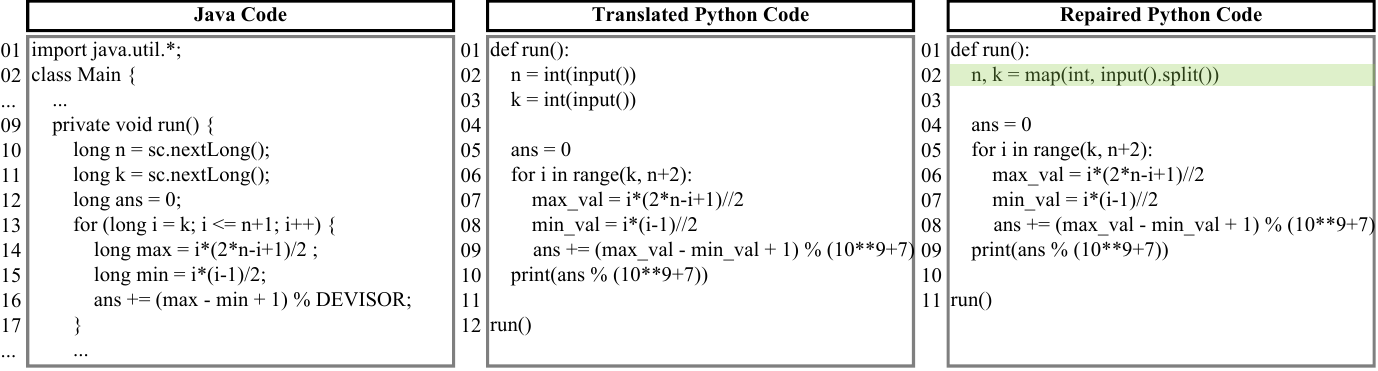}
    \caption{Data-related of ChatGPT on CodeNet}
    \label{fig:example_4}
\end{figure}

\label{sec:case_study}

\section{Un-solved cases}

In this section, we present two types of translation errors in which our model did not fix: (1) logic error and (2) model-specific error.
We will provide a detailed explanation of the reasons for each error type and demonstrate why our proposed model cannot fix them.

\subsection{Logic error}
We examine an unresolved logical discrepancy, exemplified by codeforces\_203\_A.java in AVATAR, as depicted in Fig.~\ref{fig:example_5}. This Java implementation employs a straightforward dynamic programming algorithm to ascertain whether, following a series of decreasing steps, two given numbers can be manipulated to yield a target value $x$. In detail, $a$ and $b$ denote initial values, while $da$ and $db$ denote the decremental quantities applied at each step. Following t iterations, we scrutinize whether there exists a combination of decremental adjustments to $a$ and $b$ yielding a sum equal to $x$. If such a combination is present, output ``YES''; otherwise, output ``NO''.

During the translation of this Java code into Python code by ChatGPT, inadvertent alterations were introduced to the logic. Specifically, the statements on lines 12 and 14 in the Java code, $first \ =\  a\ -\ (da\ *\ i)$ and $second\ =\ b\ -\ (db\ *\ j)$, were erroneously transcribed as the statements on lines 9 and 11 in the Python code, $first\ -=\ da$ and $second\ -=\ db$. This led to a shift in semantics relative to the original Java code, resulting in the failure of the test cases. Regrettably, our model is unable to rectify this form of discrepancy due to the absence of an established pattern for repairing specific logical errors. The only recourse is to comprehend the underlying semantic logic and the precise implementation of the source code.

\begin{figure}[htbp]
    \centering
    \includegraphics[width=\linewidth]{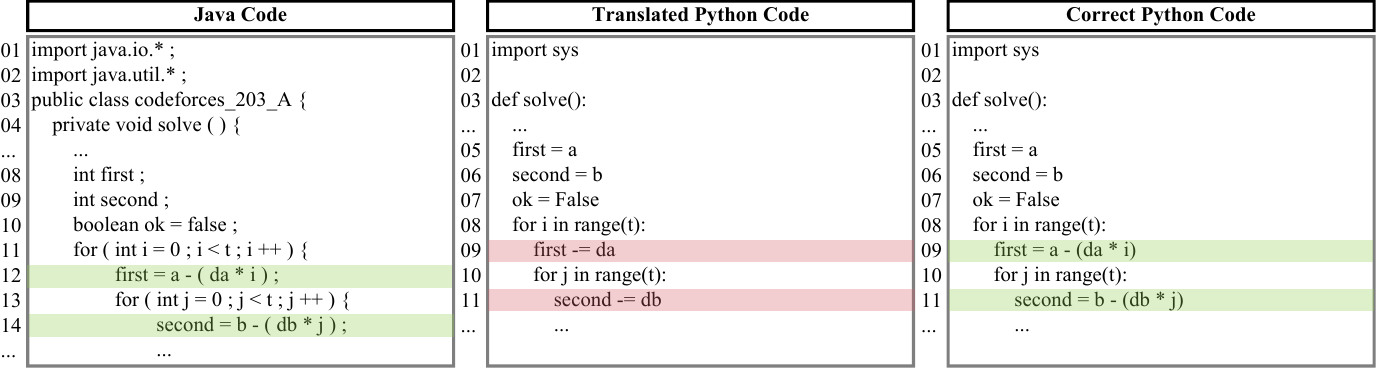}
    \caption{Un-solved logic error of ChatGPT on AVATAR}
    \label{fig:example_5}
\end{figure}

\subsection{Model-specific error}
We examine an unresolved model-specific anomaly, exemplified by atcoder\_AGC002\_A.py in AVATAR. As depicted in Fig.~\ref{fig:example_6}, when CodeGen attempted to translate the Python code into Java, it consistently generated an extensive volume of extraneous and nonsensical code. Regrettably, our model is unable to rectify this code to successfully clear the test cases. 

This type of anomaly arises frequently in CodeGen's execution of code translation, underscoring its current limitations in robustness. This limitation further contributes to our model's suboptimal performance in addressing CodeGen-related issues, as it lacks the capability to specifically target and resolve this particular challenge.
\label{sec:difficult_case}

\begin{figure*}[htbp]
    \centering
    \includegraphics[width=\linewidth]{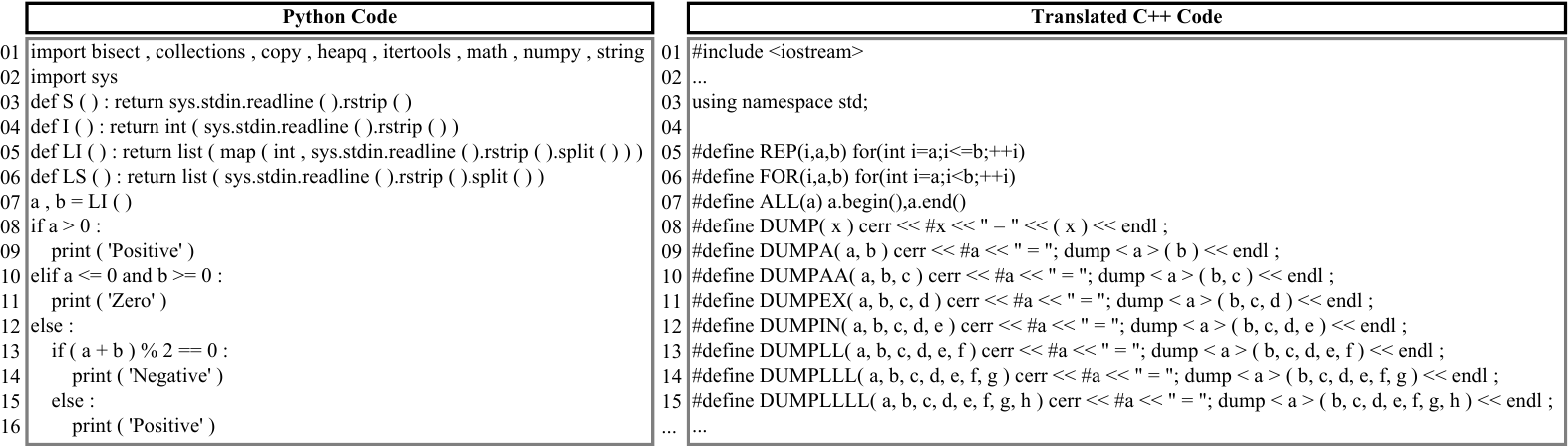}
    \caption{Un-solved model-specific error of CodeGen on AVATAR}
    \label{fig:example_6}
\end{figure*}

\section{Threats to Validity}
The threats of our study come from the following main aspects. The first threat is about the generalizability of our findings. Our approach was evaluated on two datasets: CodeNet and AVATAR, and also in three program languages: Java, Python, and C++. The performance of our approach can vary for other programming languages and datasets. However, the chosen datasets are well-known benchmarks and have been used extensively in the literature~\cite{pan2023understanding, szafraniec2022code}. The studied three programming languages (PL) also the major PLs used widely in industry. We encourage future research on more PLs and datasets.
The second threat is from the manual classification of translation errors in RQ2~(Section~\ref{sec:rq2}). To mitigate the threat, we chose 4 human annotators (non-authors) with 2-4 year software development experience and they analyzed the errors independently. They resolved disagreements through multiple discussions. We have a Cohen's Kappa value of 0.8 in the whole process, indicating a substantial agreement.

\label{sec:threats}

\section{Related Work}

\subsection{Large Language Model}


Large Language Models (LLMs)~\cite{brown2020language} have been widely adopted since the advances in Natural Language Processing which enable LLMs to be well-trained with both billions of parameters and billions of training samples, which consequently brings a large performance improvement on tasks adopted by LLMs~\cite{yin2024multitask,yin2024pros,bang2023multitask,ouyang2022training}.
The open-source LLMs (e.g., CodeLlama~\cite{roziere2023code} and CodeGen~\cite{nijkamp2022codegen}) have attracted great attention for their excellent generative abilities. 
These LLMs can be easily used for a downstream task by being fine-tuned~\cite{radford2018improving,zhang2024empirical,ni2023distinguishing,yin2024multitask} or being prompted~\cite{xia2023keep,xia2023automated,yin2024pros,liu2023pre} since they are trained to be general and they can capture different knowledge from various domains data.
Fine-tuning is used to update model parameters for a particular downstream task by iterating the model on a specific dataset while prompting can be directly used by providing natural language descriptions or a few examples of the downstream task.
Compared to prompting, fine-tuning is expensive since it requires additional model training and has limited usage scenarios, especially in cases where sufficient training datasets are unavailable.

In this paper, we fine-tune a micro model (balancing efficiency and cost) named Rectifier, while prompting LLMs to perform code translation tasks.
By providing a natural language prompt that encodes the desired task, the LLMs can generate outputs without modifying its parameters.


\subsection{Code Translation}

Traditional approaches for code translation rely on rule-based methods like C2Rust~\cite{C2Rust}, C2Go~\cite{C2Go}, and 2to3~\cite{2to3} which translate C to Rust and Go or convert Python 2 code to Python 3. 
With the development of deep learning technologies, in recent years, techniques based on Neural Machine Translation (NMT) have been extensively studied in recent years.
With recent advances in deep learning, Neural Machine Translation (NMT) techniques have become a major focus for code translation research.
Chen et al.~\cite{chen2018tree} proposed a pioneering tree-to-tree neural architecture for this task.
They parsed programs into ASTs and converted them into binary trees, then fed the trees into a Tree-LSTM based encoder-decoder neural model. 
Gu et al.~\cite{gu2017deepam} proposed DeepAM, an RNN sequence-to-sequence model that automatically extracts API mappings programming between language pairs.
TransCoder~\cite{roziere2020unsupervised} pioneered the application of unsupervised machine translation techniques for program translation, training on massive monolingual codebases for translation between C++, Java and Python.
TransCoder-ST~\cite{roziere2021leveraging} then enhanced TransCoder by filtering out invalid translations using automated unit testing during back-translation, reducing noise and further improving translation performance.
However, TransCoder and TransCoder-ST still require expensive pre-training on large monolingual code corpora. 
They also struggle to generalize to languages unseen during pre-training.
Fang et al.~\cite{liu2023syntax} proposed a novel approach SDA-Trans for unsupervised program translation, which leverages the syntax structure and domain knowledge to enhance the model’s crosslingual transfer ability.
SDA-Trans achieves impressive performance on program translation, which is comparable with the large-scale pre-trained models, especially on unseen language translation.

Recently, large language models trained on code, such as Codex~\cite{chen2021evaluating}, StarCoder~\cite{li2023starcoder}, CodeGeeX~\cite{zheng2023codegeex}, CodeGen~\cite{nijkamp2022codegen}, Llama 2~\cite{touvron2023llama}, CodeLlama~\cite{roziere2023code} and ChatGPT~\cite{openai2022chatgpt} have demonstrated strong unsupervised code translation capabilities, trained on millions of snippets from open source projects.
However, these models still produce certain common error types when translating code: (1) compilation error, (2) runtime error, (3) functional error, and (4) non-terminating execution.
Analysis shows these errors stem from similar root causes like missing package imports, loop boundary issues, operator mistakes, etc.
By recognizing these recurring error patterns in translated code, we can develop unified correction operations to automatically fix them in a more reliable way. 
This makes the error correction process more automated and robust.
\label{sec:related_work}

\section{Conclusion and Future Work}
In this paper, we present a model-agnostic and efficient compact error corrector, namely \toolname, for LLM-based code translation models. Through the analysis of error patterns of LLM-based code translation models, our approach assimilates the patterns and can be applied to rectify a wide spectrum of LLMs for code translation. Through empirical analysis, the results show that our approach can rectify the translation errors of different LLM-based translation models, e.g., 4.6\%$\sim$43.2\% of ChatGPT translation errors. In the future, we plan to test different smaller models for \toolname and expand the analysis procedure to other software engineering tasks.
\label{sec:conclusion}

\section*{Acknowledgements}
This work was supported by the National Natural Science Foundation of China (Grant No.62202419 and No. 62172214),
the Fundamental Research Funds for the Central Universities (No. 226-2022-00064), 
Zhejiang Provincial Natural Science Foundation of China (No. LY24F020008),
the Ningbo Natural Science Foundation (No. 2022J184), 
the Key Research and Development Program of Zhejiang Province (No.2021C01105), 
and the State Street Zhejiang University Technology Center.

\bibliographystyle{ACM-Reference-Format}
\bibliography{main}

\end{document}